\documentclass[conference]{IEEEtran}
\usepackage{times}

\usepackage[numbers]{natbib}
\usepackage{multicol}
\usepackage[bookmarks=true]{hyperref}
\usepackage{algorithm}
\usepackage[utf8]{inputenc} 
\usepackage[T1]{fontenc}    
\usepackage{hyperref}       
\usepackage{url}            
\usepackage{booktabs}       
\usepackage{amsfonts}       
\usepackage{nicefrac}       
\usepackage{microtype}      
\usepackage{amsmath}
\usepackage{algorithm}
\usepackage[noend]{algpseudocode}
\usepackage{algpseudocode}
\usepackage{pifont}
\usepackage{textcomp}
\usepackage{graphicx}
\usepackage{xcolor}
\usepackage{hyperref}
\hypersetup{
    colorlinks=true,
    linkcolor=blue,
    filecolor=magenta,      
    urlcolor=red,
}

\begin{document}

\title{Safe Reinforcement Learning with Mixture Density Network: A Case Study in Autonomous Highway Driving}

\author{Ali Baheri}



%
\author{\authorblockN{Ali Baheri}
\authorblockA{
West Virginia University
\\ ali.baheri@mail.wvu.edu}}

\maketitle

\begin{abstract}

This paper presents a safe reinforcement learning system for automated driving that benefits from multimodal future trajectory predictions. We propose a safety system that consists of two safety components: a heuristic safety and a learning-based safety. The heuristic safety module is based on common driving rules. On the other hand, the learning-based safety module is a data-driven safety rule that learns safety patterns from driving data. Specifically, it utilizes mixture density recurrent neural networks (MD-RNN) for multimodal future trajectory predictions to accelerate the learning progress. Our simulation results demonstrate that the proposed safety system outperforms previously reported results in terms of average reward and number of collisions.

\end{abstract}

\IEEEpeerreviewmaketitle

\section{Introduction}

The majority of research for safe automated driving has focused on rule-based approaches, inferred as handcrafted state machines. For instance, \cite{DBLP:journals/corr/abs-1708-06374} aims to formalize general requirements- called Responsibility, Sensitivity Safety (RSS), that an autonomous vehicle must satisfy for safety assurance. However, in a highly dynamic and evolving environment, there is no guarantee that rule-based approaches prevent undesirable behaviors. Furthermore, rule-based approaches are not able to generalize to unknown situations. Comparatively, fewer studies have focused on the impact of incorporating external knowledge or model for safety assurance in the learning phase. 

While model-based reinforcement learning (RL) has shown a promise in autonomous driving research \cite{baheri2020vision}, the impact of incorporating a model to address safety into the RL training phase has not been fully understood. This work is built on top of \cite{baheri2019deep,baherideep}, where a safety module for multimodal future trajectory predications is incorporated into the learning phase of RL algorithm as a model lookeahed. In contrast to the related papers, our framework benefits from the merits of both rule-based and learning-based safety approaches. Specifically, the additional safety component incorporates a multimodal future trajectory predictions model into the learning phase to predict safety longer into the future and to determine whether the future states lead to undesirable behaviors or not. If one of the future states leads to a collision, then a penalty will be assigned to the reward function to prevent collision and to reinforce to remember unsafe states. Furthermore, thanks to the nature of our state representation, the proposed framework takes into account the intentions of other road users into the decision making part. 

\begin{algorithm*}
\caption{DDQN with MD-RNN}
\begin{algorithmic}[1]
\State \textbf{Inputs:} Offline trained MD-RNN, prediction horizon $k$, number of mixture models $m$
\State \textbf{Initialize:} Safe buffer, collision buffer, $Q$-network, and target $Q$-network
\While{not done}
\State Initialize cars and obtain affordance indicators $s$
\For {length of an episode or collision}
\State Perform $\epsilon$-greedy and select action $a_t$
 \If{collision}     
 \State Reward $\leftarrow$ $R_{\mathrm{collision}}$
 \State Store $(s_t,a_t,\ast,R_{\mathrm{collision}})$ in collision buffer
 \Else 
 \State Store $(s_t,a_t,s_{t+1}, r_{t+1})$ in safe buffer
 \EndIf
 \State Use MD-RNN to predict $\big(\hat{s}^1_{t+1}, \hat{s}^1_{t+2}, \dots , \hat{s}^1_{t+k}\big)$, $\big(\hat{s}^2_{t+1}, \hat{s}^2_{t+2}, \dots, \hat{s}^2_{t+k}\big)$, $\dots$, $\big(\hat{s}^m_{t+1},  \hat{s}^m_{t+2}, \dots , \hat{s}^m_{t+k}\big)$
 \If{collision for \emph{any} future (predicted) states for \emph{any} trajectory}   
 \State Reward $\leftarrow$ $R_{\mathrm{collision}}$
 \State Store $(s_t,a_t,\hat{s}_{t+1}, R_{\mathrm{collision}})$ in collision buffer
 \EndIf
 
\State Sample random mini-batch $(s_\tau,a_\tau,s_{\tau+1},r_{\tau+1})$, $50\%$ from safe buffer and $50\%$ from collision buffer
\State Set $  y_{\tau}=\left\{
  \begin{array}{@{}ll@{}}
    r_{\tau+1} & \mbox{if sample is from collision buffer} \\
    r_{\tau+1} + \gamma \hat{Q}\Big(s_{\tau+1},\mathrm{argmax}_a \ Q(s_{\tau+1},a,\theta_\tau),\hat{\theta}_\tau\Big) & \mbox{if sample is from safe buffer}
  \end{array}\right.
$
\State Perform gradient descent on $\Big(y_\tau- Q(s_\tau,a_\tau,\theta_\tau)\Big)^2$ w.r.t $\theta$
\EndFor
\EndWhile
\end{algorithmic}
\label{alg:ddqn}
\end{algorithm*}

\section{Problem Statement and System Architecture}
\label{ref:problem statement}

\begin{figure}[t]
    \centering
    \includegraphics[width=0.45\textwidth]{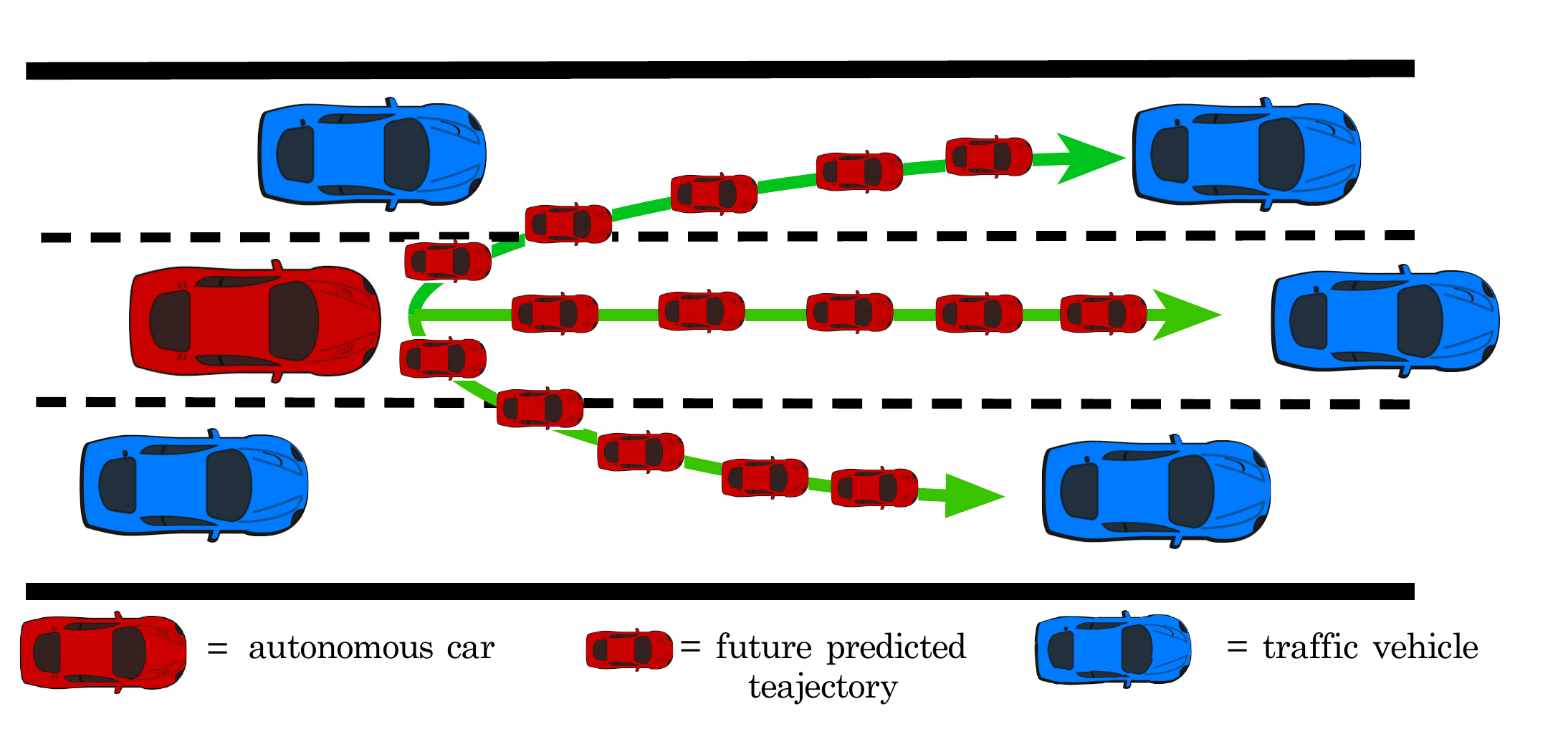}
    \caption{We study the problem of safe reinforcement learning for autonomous highway driving where the agent is capable for multimodal future trajectory predictions during the training phase.}
    \label{fig:highway}
\end{figure}

The ultimate goal of this study is to design a learning framework that is able for multimodal future trajectory predictions to address safety concerns for an autonomous vehicle in a three-lane highway scenario. Fig. \ref{fig:highway} shows this scenario. We formalize the problem as a Markov decision process (MDP) where at each time-step $t$, the agent interacts with the environment, receives the state $s_t \in \mathcal{S}$, and performs an action $a_t \in \mathcal{A}$. As a result, the agent receives a reward $r_t \in \mathcal{R}$ and ends up in a new state $s_{t+1}$. The goal is to find a policy, $\pi$, that maps each state to an action with the goal of maximizing expected cumulative reward, $\sum_{k = 0}^{\infty}\gamma^{k}r_{t+k} $, where $0 < \gamma < 1$, is the discount factor \cite{sutton1998introduction}.
The optimal action-value function, $Q^{\ast}(s,a)$, obeys the following identity known as Bellman equation,

\begin{equation}
Q^{\ast}(s,a) =\underset{{s}'} {\mathbb{E}} \ [r+ \gamma \  \underset{{a}'}\max \  Q^{\ast}({s}\textquotesingle,{a}\textquotesingle)|(s,a)].
\end{equation}
For the small scale problems, $Q^{\ast}(s,a)$, is efficiently estimated using tabular methods. For large state space problems; however, a function approximator is utilized to approximate the optimal action-value function. Approximating the optimal action-value function using a neural network associated with a few other tricks to stabilize the overall performance build the foundation of double deep Q-network (DDQN) which serves as our decision making engine in this work \cite{mnih2015human,van2016deep}.

\subsection{State space}

We assume a direct perception based approach to estimate the affordance for driving from \cite{chen2015deepdriving} for state representation. In this paper we consider a scenario where the autonomous vehicle is surrounded by up to six traffic vehicles in a three-lane highway. A total of $18$ affordance indicators are used to spatiotemporally represent the information of the six nearest traffic vehicles. These variables include the relative distance and velocity, in longitudinal direction, to the nearest front/rear car in the right/center/left lane from the autonomous vehicle\textquotesingle s perspective. 

In addition to those indicators, we use longitudinal velocity and lateral position of the autonomous vehicle. In total, these $20$ affordance indicators represent a minimal yet sufficient state representation for the three-lane highway driving scenario studied in this work.

\subsection{Action space}

We consider four action choices along longitudinal direction, namely, maintain, accelerate, brake, and hard brake. For lateral direction we assume three action choices, one for lane keep, change lane to right, and change lane to left. These result in $8$ unique action choices. 

\subsection{Reward function}

The reward function is formulated as a function of (i) desired traveling speed subject to traffic condition, (ii) desired lane and lane offset subject to traffic condition, and (iii) relative distance to the preceding car based on relative velocity as follows:

\begin{equation}
r_v = e^{-\frac{(v_{e_x} - v_{\mathrm{des}})^2}{10}} - 1,
\end{equation}
\begin{equation}
r_y = e^{-\frac{(d_{e_y} - y_{\mathrm{des}})^2}{10}} - 1,
\end{equation}
\begin{equation}
  r_x=\begin{cases}
    e^{-\frac{(d_\mathrm{lead} - d_{\mathrm{safe}})^2}{10d_\mathrm{safe}}} - 1, & \text{if $d_\mathrm{lead}<d_\mathrm{safe}$},\\
    0, & \text{otherwise.}
  \end{cases}
\end{equation}
where $v_{e_x}$, $d_{e_y}$, and $d_{\mathrm{lead}}$ are the autonomous agent\textquotesingle s velocity, lateral position, and the longitudinal distance to the lead vehicle, respectively. Similarly, ${v_\mathrm{des}}$, $y_{\mathrm{des}}$, and $d_{\mathrm{safe}}$ are the desired speed, lane position, and safe longitudinal distance to the lead traffic vehicle, respectively.

\subsection{Vehicle dynamics}

We model each vehicle using a computationally efficient point-mass model. For longitudinal equations of motion we use a discrete-time double integrator,

\begin{equation}
x(t+1) = x(t)  +v_x(t) \Delta t,
\end{equation}
\begin{equation}
v_x(t+1) = v_x(t)  +a_x(t) \Delta t,
\end{equation}
where $t$ is the time index, $\Delta t$ is the sampling time, and $x$ is the longitudinal position.  $v_x$ and $a_x$ are the longitudinal velocity and longitudinal acceleration of the vehicle, respectively. For the lateral motion, we assume a simple kinematic model,

\begin{equation}
y(t+1) = y(t)  +v_y(t) \Delta t.
\end{equation}
where $y$ is the lateral position of the car.

\section{Double Deep Q-Learning with MD-RNN}

We study the problem of safe autonomous driving for collision avoidance by introducing a learning-based model that aims to encode prior knowledge about the environment into the learning phase. The system consists of two safety components. The first module is a heuristic safety rule based on common traffic rules that ensure a minimum relative gap to a traffic vehicle based on its relative velocity, 

\begin{equation}
d_{TV} - T_{\mathrm{min}} \times v_{TV} > {d_{TV}}_{\mathrm{min}},
\label{eq:static}
\end{equation}
where $d_{TV}$, $v_{TV}$ are the relative distance and relative velocity to a traffic vehicle, $T_\mathrm{min}$ is the minimum time to collision, ${d_{TV}}_{\mathrm{min}}$ is the minimum gap which must be ensured before executing the action choice. On the other hand, the second module, predicts multimodal behavior of the future trajectories via an offline trained supervised model, detailed in Sec. \ref{sec:md-rnn}, that guides the exploration process and accelerates the learning process.

\begin{figure}[t]
    \centering
    \includegraphics[width=0.45\textwidth]{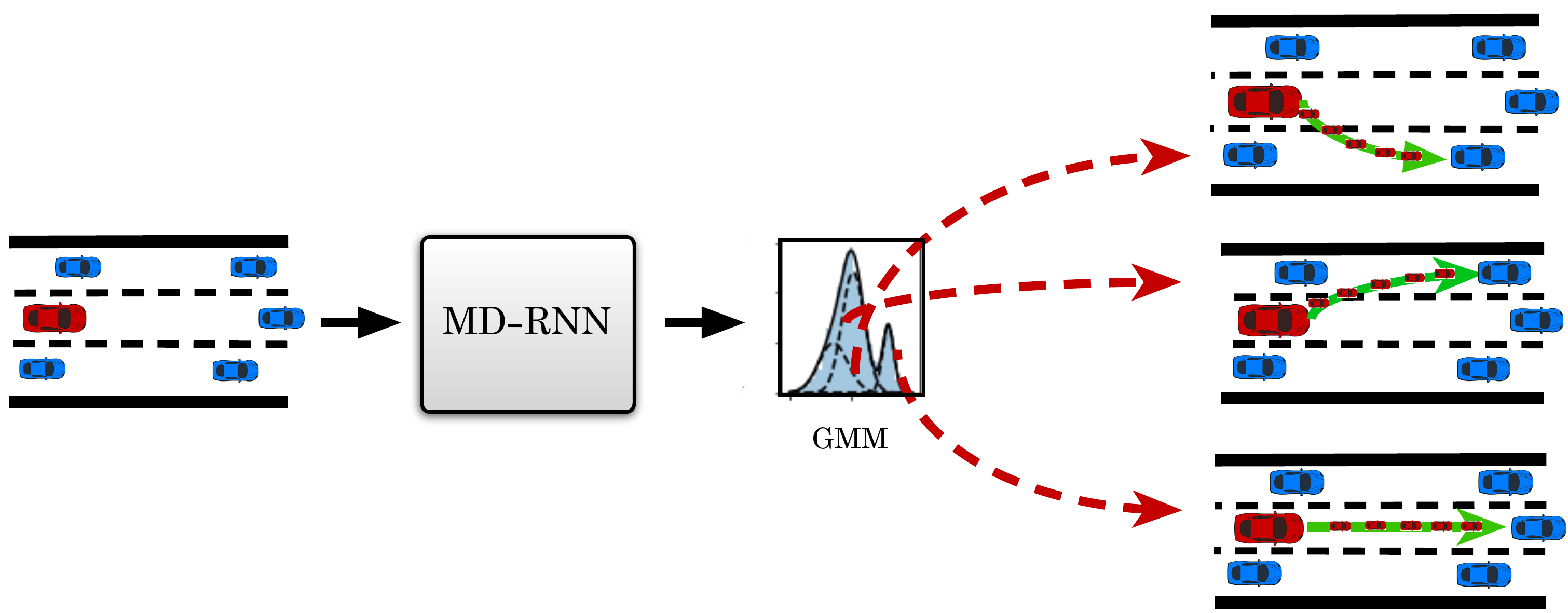}
    \caption{MDN is applied to the outputs of the RNN model that results in a Gaussian Mixture Model (GMM). MD-RNN is used for multimodal future trajectory prediction. This study considers three mixture models to predict three possible scenarios that have been shown by the green arrows.}
    \label{fig:md-rnn}
\end{figure}


\subsection{Mixture Density Recurrent Neural Networks (MD-RNN)}
\label{sec:md-rnn}

Mixture Density Networks (MDNs) \cite{bishop1994mixture} are constructed from two main components: a (recurrent) neural network and a mixture model that in principle provide a mechanism for multimodal prediction. Combined with RNN, MDN has been applied in many applications from parametric speech synthesis \cite{wang2017autoregressive} and model 2D pen data \cite{graves2013generating} to predict future state of a video game screen image to accelerate the learning process of an RL agent \cite{ha2018recurrent}.

A MDN transforms the outputs of the RNN to build the parameters of a Gaussian mixture model (GMM) that is a convex combination of Gaussians. A GMM is the weighted sum of many Gaussians with different means and standard deviations. The central idea of a MDN is to predict an entire probability distribution of the output(s) instead of generating a single prediction. We apply the MDN to the outputs of an RNN to predict the future trajectories of the autonomous vehicle. In our previous work, we trained an RNN whose inputs were the pair of states and action and the output was a single future trajectory of the agent. In contrast to our previous work, the MD-RNN outputs a GMM for multimodal future trajectory predictions that each mixture component describes a certain driving behavior (Fig. \ref{fig:md-rnn}). 

In \cite{baheri2019deep} we demonstrated that predicting a single trajectory of the agent consisting of future states in a given horizon, checking if one of the future states leads to a collision, and incorporating this knowledge into the training phase of RL accelerate the learning process and significantly reduces the number of collisions. The present study extends our previous work by incorporating multimodal future trajectory predictions that represent different driving behaviors using the MDN. 

To collect data for MD-RNN training, we train an RL agent without a learning-based safety module and collect a long history of states and corresponding action that builds our driving data. Once MD-RNN trained, we check whether future states lead to an accident within a pre-defined finite horizon. If one of the future states for any predicted trajectory leads to an accident, we assign a negative reward to remember unsafe states and accelerate the learning process. 

We summarize the DDQN with MD-RNN in Algorithm \ref{alg:ddqn}. The algorithm is initialized with two buffers, namely the safe and collision buffer, to store good and bad behaviors. At each time step, we check whether the immediate collision occurs using Eq. \ref{eq:static}. If not, we store it in the safe buffer. Otherwise, we store the danger state in the collision buffer and assign a large negative reward, $R_\mathrm{collision}$, to the reward function. Next, we use the MD-RNN for multimodal trajectory predictions for a given horizon and check whether there exists any violation of safety rules (Eq. \ref{eq:static}). In case of violation, we store the next state in the collision buffer and assign a large negative reward, $R_\mathrm{collision}$, to the reward function. To update the temporal difference target, we equally sample from both collision and safe buffers. Finally, the model parameters are updated using a stochastic optimization algorithm.

\section{Results}

\begin{figure}[t]
    \centering
    \includegraphics[width=0.5\textwidth]{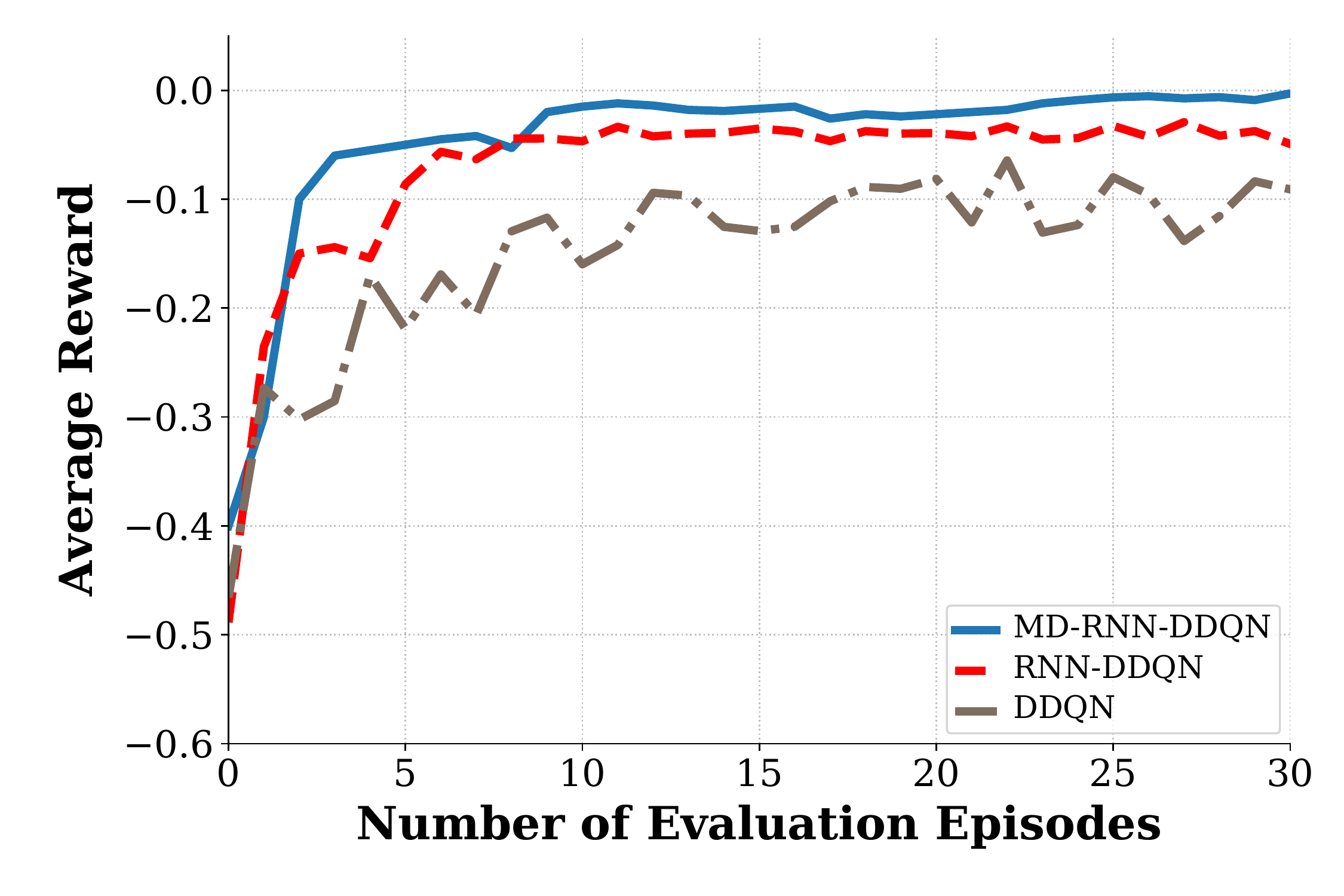}
    \caption{Learning curves during training. We train three policies (i) DDQN without RNN (ii) DDQN with RNN and (iii) DDQN with MD-RRN, for a total of $3000$ episodes. We evaluate these policies every $100^{\mathrm{th}}$ episode and report the average cumulative reward.}
    \label{fig:learning_curve}
\end{figure}

\begin{figure}[t]
    \centering
    \includegraphics[width=0.5\textwidth]{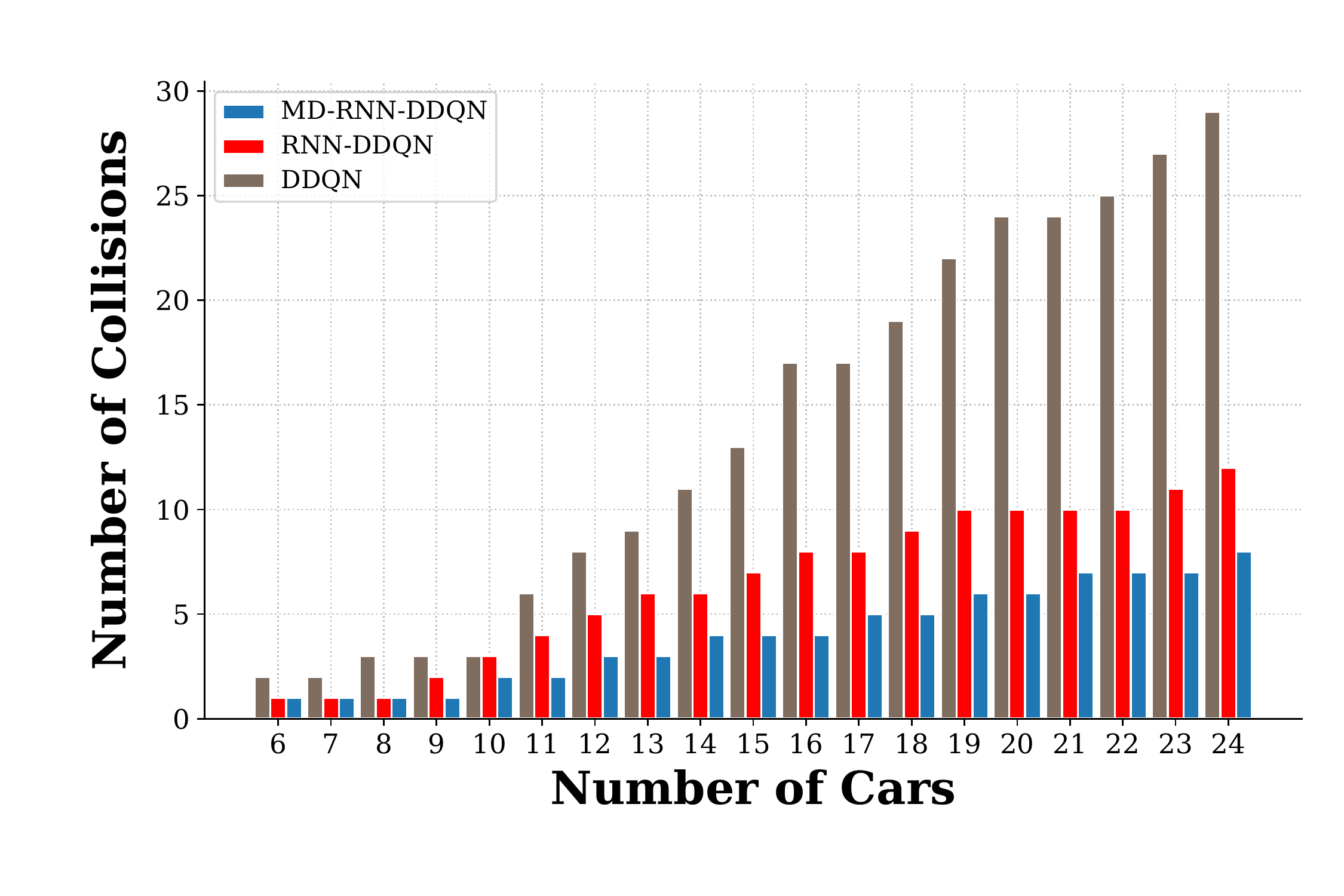}
    \caption{Number of collisions for different numbers of traffic vehicles after training. We evaluate three policies (i) DDQN without RNN (ii) DDQN with RNN and (iii) DDQN with MD-RRN after training for $3000$ times.}
    \label{fig:collision}
\end{figure}

We evaluate the effectiveness of the proposed framework in a simulation environment. The autonomous agent utilizes $\epsilon$-greedy strategy to make decisions. Other vehicles are controlled externally. Furthermore, other system parameters such as maximum velocity are randomly chosen for traffic vehicles. We train our autonomous agent for a total of 3000 episodes. Each episode is initialized with randomly chosen different number of vehicles. Each episode terminates when the autonomous agent collides with a vehicle, or when a time budget is exhausted. During the learning phase, we partially evaluate the proposed architecture every episode. Fig. \ref{fig:learning_curve} represents the cumulative reward during the training phase. It can be seen that the policy with MD-RNN model outperforms the policy with and without RNN model. We also evaluate two policies after training for $3000$ times for different number of vehicles ranging from $6$ to $24$. Fig. \ref{fig:collision} demonstrates that as the number of vehicles increases, the number of collisions increases, as expected.

\section{Conclusion}
We proposed a reinforcement learning architecture for safe automated driving in a three-lane highway scenario that utilizes a multimodal trajectory predictions. This model was served as a model lookahead to accelerate the learning process and guid the exploration process. We argued that heuristic safety rules are susceptible to deal with unexpected behaviors particularly in a highly changing environment. To alleviate this issue, we proposed a learning-based mechanism to learn safety patterns from driving data. To achieve that goal, we trained a mixture density recurrent neural network (MD-RNN) to predict a set of future trajectories and determine whether one of the future sates in any of these trajectories violates the safety rule. We demonstrated that incorporating this knowledge into the training phase accelerates the learning process and results in significantly less collisions. 

\bibliographystyle{unsrt}
\bibliography{BOSummer2016}

\end{document}